\documentclass[reprint,
superscriptaddress,
amsmath,
amssymb,
aps,
prb,
floatfix,
english
]{revtex4-2}

\usepackage{bibunits}
\usepackage[unicode=true, colorlinks=true, citecolor={blue!80!black}, urlcolor={blue!50!black}, linkcolor = {blue!80!black}]{hyperref}
\defaultbibliography{references}
\defaultbibliographystyle{apsrev4-2}
\usepackage{graphicx}
\usepackage{svg}
\usepackage{wasysym}
\usepackage{dcolumn}
\usepackage{siunitx}
\usepackage[T1]{fontenc}
\usepackage[utf8]{inputenc}
\usepackage{bbm}

\usepackage{sidecap}
\sidecaptionvpos{figure}{t}

\begin{document}
\widetext

\title{Bloch oscillations in a transmon embedded in a resonant electromagnetic environment}
\author{Benjamin Remez}
\email{benjamin.remez@yale.edu}
\affiliation{Department of Physics, Yale University, New Haven, CT 06520, USA}
\author{Vladislav D.~Kurilovich}\thanks{
Present address: Google Research, Mountain View, CA, USA.}
\affiliation{Department of Physics, Yale University, New Haven, CT 06520, USA}
\author{Maximilian Rieger}
\affiliation{Department of Physics, Yale University, New Haven, CT 06520, USA}
\author{Leonid I.~Glazman}
\affiliation{Department of Physics, Yale University, New Haven, CT 06520, USA}

\begin{abstract}
    Recently developed Josephson junction array transmission lines implement strong-coupling circuit electrodynamics compatible with a range of superconducting quantum devices.
    They provide both the high impedance which allows for strong quantum fluctuations, and photon modes with which to probe a quantum device, such as a small Josephson junction.
    In this high-impedance environment, current through the junction is accompanied by charge Bloch oscillations analogous to those in crystalline  systems. However, the interplay between Bloch oscillations and environmental photon resonances remains largely unexplored. 
    Here we describe the Bloch oscillations in a transmon-type qubit attached to high-impedance transmission lines with discrete photon spectra. Transmons are characterized by well-separated charge bands, favoring Bloch oscillations over  Landau-Zener tunneling. We find resonances in the voltage--current relation and the spectrum of photons emitted by the Bloch oscillations.
    The transmon  also scatters photons inelastically; we find the cross-section for a novel anti-Stokes-like process whereby photons gain a Bloch oscillation quantum.  Our results outline how Bloch oscillations leave fingerprints for experiments across the DC, MHz, and GHz ranges.
\end{abstract}

\maketitle

\section{Introduction}
The introduction of the fluxonium qubit \cite{Manucharyan2009, Nguyen2019,Somoroff2023} touched off the development of high-impedance, low-loss circuit elements, dubbed superinductors \cite{Manucharyan2012,Bell2012, Masluk2012,Grunhaupt2019, Niepce2019, Peruzzo2020, Rieger2023}. By now, their use extended beyond the superconducting devices of quantum information technology. Superinductors are used to engineer a high-impedance electromagnetic environment crucial for allowing strong quantum phase fluctuations in a nominally-superconducting circuit. The simplest circuit of this kind consists of a small-capacitance Josephson junction shunted by a superinductor \cite{Kuzmin2019}. Quantum fluctuations of phase across  the small junction were demonstrated by observing of inelastic photon scattering with the production of a number of smaller-frequency microwave photons \cite{Burshtein2021,Mehta2023}. A more recent microwave experiment \cite{Kuzmin2023} provided evidence in favor of the Schmid dissipative quantum phase transition \cite{Schmid1983}. This transition from a superconducting to insulating state occurs once the zero-frequency impedance $Z\equiv Z(\omega\to 0)$ of the superinductor reaches a critical value, the resistance quantum $R_Q=h/(4e^2)$. 

In the insulating state ($Z>R_Q$), Coulomb blockade of the small junction wins over the Josephson effect. The resulting differential resistance of the junction, $R(I)=dV/dI$, becomes a decreasing function of the direct current (DC) passing through the circuit. The dissipative nature of the DC charge transport is associated with the Bloch oscillation  of charge occupying the small junction capacitance \cite{Likharev1985, Kuzmin1991,  Watanabe2001}. These oscillations radiate waves propagating along the superinductor and carrying energy away from the junction. The  Bloch oscillations frequency is determined by the DC current and the Cooper pair charge, $\Omega/(2\pi)=I/2e$.

The notion of the Schmid transition and aforementioned dissipation mechanism assume a featureless frequency dependence  of the superinductor impedance $Z(\omega)$ ``seen'' by the small junction. A featureless dependence, approaching a finite constant $Z$ in the limit $\omega\to 0$, would require an infinitely-long unrealistic superinductor.

A superinductor of finite length, along with the impedance mismatch at its interface with the external microwave circuit, results in standing wave resonances and a frequency-dependent $Z(\omega)$. Unless carefully matched \cite[see, e.g.,][]{Altimiras2013, Planat2020},  the impedance mismatch is typically strong \cite{Leger2019,PuertasMartinez2019, Kuzmin2019} and the resonances are narrow. The effect of resonances in $Z(\omega)$ on the  photon inelastic scattering off a small junction is well documented \cite{Mehta2023}. Nevertheless, how they pertain to Bloch oscillations and the Schmid picture is not yet clear. 

Our goal is thus two-fold. First, to find the ramifications narrow resonances in $Z(\omega)$ have for the $V(I)$ dependence in the nominally-insulating small junction.
Second, to build a theory of photon emission and inelastic photon scattering in the presence of Bloch oscillations excited by a direct current passing through the circuit. 
We focus on the most interesting case of high-impedance Josephson junctions arrays ($Z \gg R_Q$), allowing for strong Bloch oscillations, and transmon junctions, for which the insulating state is the most striking. 
In the following Sections, we introduce the model for a DC-biased transmon (Sec.~\ref{sec:model}), analyze the manifestations of Bloch oscillations in the voltage-current characteristic (Sec.~\ref{sec:voltage}), in photon radiation (Sec.~\ref{sec:radiation}), and in  inelastic photon scattering (Sec.~\ref{sec:upmixing}), and draw connections to present experimental platforms (Sec.~\ref{sec:conclusions}). All the while, we will bring the effects of impedance mismatch to the fore.

\begin{figure}
    \centering
    \includegraphics[width = 8.6cm]{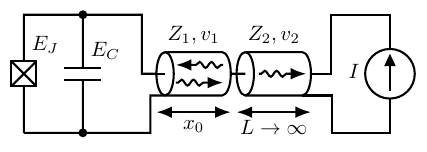}
    \caption{The circuit under consideration. A transmon qubit is shunted by a Josephson junction array superinductor of length $x_0$, which at low frequencies behaves as a waveguide with high impedance $Z_1 (0)$. The waveguide is contacted by  microwave measurement circuitry with low impedance $Z_2(0)$. External DC bias $I$ drives qubit Bloch oscillations, which excite photons in the array (wavy lines). Reflection at the impedance-mismatched interface leads to standing wave modes $\omega_n$.}
    \label{fig:fig1}
\end{figure}

\section{Low-frequency circuit model}
\label{sec:model}

We consider the circuit depicted in Fig.~\ref{fig:fig1},  of a small Josephson junction terminating a transmission line, also comprised of Josephson junctions \cite{Manucharyan2012, Kuzmin2022}. The Hamiltonian of an isolated junction is
\begin{equation}\label{eq:junction}
    H_J = 4E_C (\hat N - n)^2 - E_J \cos \hat \varphi.
\end{equation}
The phase difference across the junction, $\hat\varphi$, is a canonical conjugate to the number of transferred Cooper pairs $\hat N$: $[\hat\varphi,\hat N ]~=~i$. $E_C$ and $E_J$ are the junction charging and Josephson energies, respectively, and $n$ is the displacement charge across the junction (see below).
In a transmon \cite{Koch2007}, the Josephson plasma frequency $\omega^{\phantom\dagger}_{\rm Q}=\sqrt{8E_JE_C}/\hbar$ substantially exceeds the widths $\lambda_m$ of the first few ``bands'' modulated by the displacement charge $n$.
This is achieved by making the ratio $E_J/E_C$ sufficiently large. The same condition establishes a hierarchy between the charge bandwidths, 
\begin{equation}\label{eq:hierarchy}
\omega^{\phantom\dagger}_{\rm Q}\gg |\lambda_1| \gg |\lambda_0|,
\end{equation}
involving the two lowest bands \cite{Koch2007}
\begin{eqnarray}
    &&\lambda_0 = -E_C\,2^5\sqrt{\frac{2}{\pi}} \Bigl(\frac{E_J}{2E_C}\Bigr)^{3/4} e^{-\sqrt{8E_J / E_C}},\nonumber\\
    &&\lambda_1=-8\sqrt{\frac{E_J}{E_C}}\lambda_0\,
    \label{eq:lambdas}
\end{eqnarray}
(hereinafter we use units with $\hbar=1$). The bandwidths $\lambda_{0,1}$ are associated with the amplitude of a phase slip across the Josephson junction in the absence or presence of a plasmon excitation, respectively. Hence, the energy separation between the bands is approximately $\hbar\omega^{\phantom\dagger}_{\rm Q}$. 

Next, consider the charge waves propagating along the Josephson junction array. We assume its junctions belong to the ``classical'' limit, corresponding to a negligible probability for their phases to slip. Therefore, the array serves as a harmonic medium for the waves. Their spectrum $\omega(k)$ is linear, $\omega(k)=v_1k$, for wavelengths longer than the unit cell $a$ of the periodic array, $k\ll 2\pi/a$. [In practice, the junctions' plasma resonance places a lower cutoff on $k$.] The excitations in the linear part of the spectrum can be described \cite{Fisher1997} in terms of canonically conjugated fields of phase $\phi(x)$ and charge density $\rho(x)=2e\partial_x\theta/\pi$, 
\begin{equation}\label{eq:H_array}
    H_{\rm array} = \int_0^{x_0} dx\,\frac{v_1}{2\pi}\left[\frac{1}{K}(\partial_{x}\theta)^{2}+K(\partial_{x}\phi)^{2}\right].
\end{equation}
Here $v_1$ is the wave velocity in the array. Fields $\theta(x)$ and $\phi(x)$ satisfy the following commutation relation $[\partial_x \theta(x), \phi(x^\prime)] = i\pi \delta (x - x^\prime)$. Finally, $K$ is a dimensionless parameter characterizing the wave impedance of the array:
\begin{equation}
    K = \frac{R_Q}{2Z_1(0)},\quad\quad R_Q = \frac{h}{4e^2}\,.
\end{equation}
We will consider only circuits deep on  the insulating side of the Schmid transition, that is $K\ll1/2$,  which occurs at fairly high impedance $Z_1$. To capture the impedance mismatch at the interface with the external low-impedance circuit, we complement Eq.~(\ref{eq:H_array}) with Hamiltonian
\begin{equation}\label{eq:H_envrnmnt}
    H_{\rm low-imp} = \int_{x_0}^L dx\,\frac{v_2}{2\pi}\left[\frac{1}{K_{\rm e}}(\partial_{x}\theta)^{2}+K_{\rm e}(\partial_{x}\phi)^{2}\right],
\end{equation}
and a proper current conservation condition at $x=x_0$. The dimensionless parameter $K_{\rm e}=R_Q/2Z_2(0)$ corresponds to the low-frequency impedance of the environment.

Taking the lead from experiments \cite{Kuzmin2019a,Kuzmin2021, Mehta2023}, we assume that the transmon transition frequency $\omega_Q$ falls into the linear part of the spectrum $\omega(k)$. That allows us to use Eq.~(\ref{eq:H_array}) in the description of the transmon galvanically coupled to the transmission line. Once coupled, charge $n$ is promoted to operator $\hat n$,
\begin{equation}
    \hat n =  {\cal N}- \theta(x = 0) / \pi.
\end{equation}
The first contribution here is a $c$-number that describes the external direct current bias $I$, $d{{\cal N}}/dt = I/ 2e$. The second contribution describes dynamical  charge (in units of $2e$) associated with quantum fluctuations in the line.
The galvanic coupling with the array stipulates $\varphi=\phi(x=0)$.

In what follows, we will focus on the range of current values $I$ satisfying 
\begin{equation}\label{eq:current_limits}
\omega_{\rm Q} \gg \pi I/e.
\end{equation}
This condition means the  oscillations in $\hat n$ induced by the current are much slower than the transmon's Josephson plasma frequency.
Therefore, a transmon initially in its ground state remains in the lowest Josephson plasmon ``band'' (i.e., in the ground state $|0\rangle$ defined by the instantaneous displacement charge $\hat n$) upon application of the current $I$.
 By projecting Eq.~\eqref{eq:junction} onto this band, we find  
\begin{equation}\label{eq:junction_0}
    H_{|0\rangle} = - \lambda_0 \cos(2\pi \hat n)\equiv -\lambda_0\cos[2\theta(x = 0) - \pi It/e].
\end{equation}
The sinusoidal modulation in Eq.~(\ref{eq:junction_0}) is the putative Bloch oscillation, occurring with frequency
\begin{equation}\label{eq:bloch_freq}
    \Omega = \pi I /e.
\end{equation}

The junction nonlinearity causes Bloch oscillations to also modulate  the scattering of high-frequency ($\omega\approx\omega_Q^{\phantom\dagger}$) photons {off the transmon}. 
This is analogous to the mixing of  a baseband signal (Bloch oscillation) and a carrier wave (photon)  in radio transceivers. 
The mode mixing occurs due to the low-frequency modulation of the energy of the excited transmon state $|1\rangle$ described by a projected Hamiltonian
\begin{equation}\label{eq:junction_1}
    H_{|1\rangle} = -\lambda_1 \cos(2\pi \hat n)\equiv -\lambda_1\cos[2\theta(x = 0) - \pi It/e].
\end{equation}
Condition $|\lambda_1|\gg |\lambda_0|$, see Eq.~(\ref{eq:hierarchy}), allows us, in the context of mode mixing in 
Sec.~\ref{sec:upmixing}, to disregard the charge modulation of the ground-state energy. Therefore, we may proceed by adapting the formalism developed in Ref.~\cite{Houzet2020} to quantify the sideband formation in the photon scattering process.

\section{Voltage--current relation in the presence of impedance mismatch}
\label{sec:voltage}
In typical implementations, there is a large impedance mismatch between the two segments described by Eqs.~(\ref{eq:H_envrnmnt}) and (\ref{eq:H_array}), respectively: $Z_2(0) = 50\,\Omega$ while $Z_1(0)$ can be as high as $20\,{\rm k}\Omega$.
The impedance mismatch results in the reflection of waves off the interface between the segments at $x = x_0$. The plasmon reflection amplitude is
\begin{equation}
    r = \frac{Z_1(0) - Z_2(0)}{Z_1(0) + Z_2(0)}.
\end{equation}  
In what follows, we assume that $Z_1 (0) > Z_2 (0)$, i.e., $r > 0$.  Reflection leads to the formation of standing waves in the array. This results in a resonant structure in the $VI$-relation of the transmon. We show that an appreciable voltage develops across the transmon only when the Bloch oscillation frequency $\Omega = \pi I / e$
(determined by the bias current $I$) is resonant with one of the standing wave frequencies.

The coupling between the Josephson junction and the transmission line is described by Eq.~\eqref{eq:junction_0}. To elucidate the effect of the coupling on the charge dynamics, we expand the displacement operator $\theta(x = 0)$ into the photon modes in the line. The expansion reads
\begin{equation}\label{eq:theta}
    \theta (x=0) = \sum_q \sqrt{\frac{K \Delta}{\omega_q}}\,\theta_q\,\bigl(a_q + a_q^\dagger\bigr).
\end{equation}  
Here $a_q^\dagger$ and $a_q$ are the creation and annihilation operators of excitations propagating in the entire electromagnetic environment   described by Hamiltonian
\begin{equation}\label{eq:H0}
H_0\equiv H_{\rm array}+ H_{\rm low-imp}=\sum_q\omega_q a_q^\dagger a_q^{\vphantom \dagger},
\end{equation}
see Eqs.~(\ref{eq:H_array}) and (\ref{eq:H_envrnmnt}). Parameter $\Delta = \pi v_2 / L$ is the spacing between two subsequent modes supported by the whole line, where $L\rightarrow \infty$ is the normalization length. In that limit, the mode spectrum is a continuum $\omega_q=v_2 q$, with $q$ the mode wavenumber in the long external segment \cite{Indices}. We denote the displacement field in a mode $q$ by $\theta_q(x)$. The impedance mismatch is reflected in the dependence of $\theta_q \equiv \theta_q(x = 0)$ on $\omega_q$:
\begin{equation}\label{eq:theta_q_mis}
    \theta_q = \sqrt{\frac{1 - r^2}{(1 - r)^2 + 2 r (1 - \cos \omega_q t_0)}},
\end{equation}
where $t_0 = 2x_0 / v_1$ is the round trip time for a wave in the array ($0 < x < x_0$). 
For $1 - r \ll 1$, $\theta_q$ has sharp resonance peaks at standing wave frequencies $\omega_q = n\cdot \delta\omega$, where $n$ in a non-negative integer and
\begin{equation}\label{eq:deltaomega}
    \delta\omega = \frac{2\pi}{t_0}=\frac{\pi v_1}{x_0}
\end{equation}
is the spacing between the resonant modes.

We find the $VI$-relation in two steps. First, we compute the power $P(I)$ dissipated from the junction at a given current $I$ in the form of waves. Then, division by $I$ yields the dissipative DC voltage as a function of current, $V(I)=P(I)/I$. 
To evaluate the power $P(I)$ dissipated into the electromagnetic environment, we apply Fermi's golden rule. Denoting the initial and final states of the system as $|i\rangle$ and $|f\rangle$, respectively, we obtain at $T = 0$:
\begin{align}\label{eq:P_start}
    P(I) = 2\pi \frac{\pi I}{e} \sum_f |\langle f|\lambda_0\, e^{2i\theta} / 2|i\rangle|^2 \delta(E_f - E_i - \pi I/e),
\end{align}
where we abbreviated 
$\theta \equiv \theta(x = 0)$. Performing the summation over the final states, we find
\begin{align}
    &P(I) = \frac{\lambda_0^2}{4}\frac{\pi I}{e}{\cal C}_\theta(\Omega=\pi I/e),\,\,\,{\cal C}_\theta(\Omega)=
    \int dt e^{-i\Omega t}{\cal C}_\theta(t),\notag
    \\
    &{\cal C}_\theta(t) = \langle e^{-2i\theta^{(0)}(0)} e^{2i\theta^{(0)}(t)}\rangle.\label{eq:P_to_C}
\end{align}
Here the averaging is over the ground state of the system unperturbed by Eq.~\eqref{eq:junction_0} and $\theta^{(0)}(t) = e^{iH_0 t} \theta e^{-iH_0 t}$ is the boundary displacement operator in the interaction picture. To evaluate  the charge correlation function  ${\cal C}_\theta(t)$, we use the mode expansions, Eqs.~(\ref{eq:theta}) and (\ref{eq:H0}), which allows us to factorize ${\cal C}_\theta(t)$ into a product over modes. The ground-state average for a single mode is a Gaussian function of $\theta_q$, see Eq.~(\ref{eq:theta_q_mis}). Combining contributions of different modes, we obtain for the correlation function
\begin{equation}\label{eq:C_to_J}
    {\cal C}_\theta(t) = e^{-J(t)},
\end{equation}
where
\begin{equation}\label{eq:J_mismatch}
    J(t) = 4K\int_0^{+\infty}\frac{d\omega}{\omega}\frac{(1 - r^2)\bigl(1-e^{i\omega t}\bigr)e^{-\omega / \omega_{\rm Q}}}{(1-r)^2 + 2r(1 - \cos \omega t_0)}.
\end{equation}
Here we regularized the logarithmically divergent integral by introducing a factor $e^{-\omega / \omega_{\rm Q}}$; the cutoff frequency scale~\cite{Houzet2023} is of the order of the transmon plasma frequency $\omega_{\rm Q} = \sqrt{8 E_J E_C}$.

Equations \eqref{eq:P_to_C}--\eqref{eq:J_mismatch} apply to any impedance mismatch. For instance, for matched impedances $r~=~0$ and we reproduce the known \cite{Weiss1985, Weiss1987} result for ${\cal C}_\theta(\Omega)$. Typical experimental parameters correspond to the regime of {\it strong} impedance mismatch; in it, the reflection amplitude is close to unity, $1 - r \sim 0.01 \ll 1$. From now on, we focus on this practically relevant case.
There it is useful to introduce the loss rate for the modes in the array:
\begin{equation}
    \gamma = \frac{1 - r^2}{2t_0}=\frac{\delta\omega}{\pi}\frac{Z_1(0)Z_2(0)}{[Z_1(0)+Z_2(0)]^2}.
\end{equation}
(more precisely, $\gamma$ is a contribution to the loss rate associated with the mode leakage into the low-impedance circuit). For $Z_1(0) \gg Z_2(0)$, the loss rate satisfies the condition $\pi\gamma \ll \delta \omega$. That allows us to represent the part of the integrand in Eq.~\eqref{eq:J_mismatch} periodic in $\omega$ as a sum over Lorentzian resonances:
\begin{align} \label{eq:thetaq_lorentzians}
    \frac{1 - r^2}{(1-r)^2 + 2r(1 - \cos \omega t_0)}\approx \sum_{n = 0}^{+\infty} \frac{\delta \omega}{2\pi}\frac{ 2\gamma}{\gamma^2 + (\omega - \omega_n)^2},
\end{align}
where $\omega_n = n\cdot \delta \omega$ \cite{Indices}. Substituting this representation into Eq.~\eqref{eq:J_mismatch}, we separate $J(t)$ into the form
\begin{equation}\label{eq:J_total}
    J(t) = J_0(t) + J_{\rm res}(t).
\end{equation}
Here $J_{\rm res}(t)$ encodes the contributions of modes with $n \geq 1$, while $J_0(t)$ comes from $n = 0$ only.  For  $J_{\rm res}(t)$ we find
\begin{equation}\label{eq:J_d}
    J_{\rm res}(t) = 4K\sum_{n = 1}^{+\infty}\frac{\delta \omega}{\omega_n}\bigl(1 - e^{-\gamma |t|}e^{i\omega_n t}\bigr)e^{-\omega_n / \omega_{\rm Q}}.
\end{equation}
The $n=0$ contribution can be expressed in terms of a dimensionless function $\mathcal{J}(\gamma t)$:
\begin{align}\label{eq:soft_J}
    J_0(t) = \frac{4K\delta \omega}{\gamma} \mathcal{J}(\gamma t),\quad \mathcal{J}(y) = \frac{1}{\pi} \! \int_0^{\infty} \! \frac{dx}{x} \frac{1-e^{iyx}}{1 + x^2}. 
\end{align}
$J_0(t)$  varies on a ``slow'' timescale $\sim 1/\gamma$. Its influence on the resulting resonant structure of $P(I)$, Eq.~(\ref{eq:P_to_C}), is inconsequential, somewhat affecting only the shape of well-separated resonances. Therefore, in the following we dispense with the $J_0(t)$ contribution \cite{J0}.

We now apply Eqs.~\eqref{eq:J_total}--\eqref{eq:soft_J} in conjunction with Eqs.~\eqref{eq:P_to_C} and \eqref{eq:C_to_J} to find $P(I)$ and $V(I)$ under strong impedance mismatch. Having neglected $J_0(t)$, we can cast Eq.~\eqref{eq:P_to_C} in the following form:
\begin{equation}\label{eq:P_in_modes}
    P(I) = \frac{\pi\lambda_0^2}{4e} \Bigl(\frac{\delta \omega}{\omega_{\rm Q}}\Bigr)^{4K} I
    \int dt e^{-i\pi It/e} \prod_{n=1}^{+\infty}e^{4K e^{i\omega_n t} e^{-\gamma |t|} / n }.
\end{equation}
Factor $(\delta \omega / \omega_{\rm Q})^{4K}$ describes the renormalization of the amplitude of phase slips by the high-frequency modes; it stems from the first term in parentheses in Eq.~\eqref{eq:J_d}. Exponential factor $e^{4K e^{i\omega_n t} e^{-\gamma |t|} / n}$ accounts for the emission of photons of frequency $\approx \omega_n$. A $p$-th term in the series expansion of this factor,
\begin{equation}\label{eq:series}
    e^{4K e^{i\omega_n t} e^{-\gamma |t|} / n} = \sum_{p=0}^{+\infty} \frac{(4K/n)^p}{p!}e^{ip\omega_n t} e^{-p\gamma |t|},
\end{equation}
describes a process with an emission of $p$ photons of frequency $\omega_n$. Application of expansion \eqref{eq:series} to each exponent in Eq.~\eqref{eq:P_in_modes} breaks down $P$ into the contributions of individual multiphoton processes. The simplest process is the one in which driving by current $I$ 
excites a single photon with frequency \begin{equation}\label{eq:bloch_freq}
    \Omega=\pi I/e.
\end{equation}
This is the nominal Bloch oscillation frequency. Were the single-photon emission the only process possible, we would find in $P(I)$ resonant lines of a fixed, small width $e\gamma/\pi\ll(e/\pi)\delta\omega$ at $I \approx e\omega_N/\pi$ (here $N=1,2,3,\dots$).
However, a photon with frequency $\omega_N$ can also ``split'' into several photons of lower frequency, $\omega_N = \sum_{i = 0}^{q} \omega_{n_i}$. As we will show, the higher the $N$, the larger the typical number of emitted photons, $m_{\rm typ}$.
The fixed ``uncertainty'' $\gamma$ in the frequency of each additional photon results in the linewidths increasing  with the  resonance mode number $N$. For sufficiently high $N$, the resonances are washed out completely.

The phase space available for the photon splitting is vast when $N \gg 1$. Therefore, the frequencies of photons produced in a splitting event are almost always pairwise different, motivating us to restrict the summation in Eq.~\eqref{eq:series} to $p = 0$ and $1$ \footnote{This approximation is justified under the made assumption $K\ll 1$.} . Using this approximation, we can represent $P(I)$ as 
\begin{widetext}
\begin{equation}\label{eq:P_sums}
    P(I) = \frac{\pi\lambda_0^2}{4e}I\Bigl(\frac{\delta \omega}{\omega_{\rm Q}}\Bigr)^{4K}\sum_N \int dt e^{-i(\pi I/e - \omega_N)t}\sum_{m = 1}^N \frac{(4K)^m}{m!}\sum_{n_1,\dots,n_m} \frac{\delta_{n_1 + \dots + n_m, N}}{n_1\cdot \dots \cdot n_m} e^{-\gamma m |t|}.
\end{equation}
Here $q$ is the total number of the emitted photons in which the drive  quantum of frequency $\pi I/e$ splits. The sum over $n_{i}$ depends logarithmically on $N$. For $N\gg 1$, we find
\begin{equation}\label{eq:logs}
\sum_{n_1,\dots,n_q} \frac{\delta_{n_1 + \dots + n_m, N}}{n_1\cdot \dots \cdot n_m} \sim \frac{m \ln^{m-1}N}{N}.
\end{equation}
Using this relation in Eq.~\eqref{eq:P_sums}, noting that at given current $I$ the relevant $N$ is $N \approx \pi I/ e\delta \omega$, and rearranging the terms, we obtain
\begin{equation}\label{eq:P_sums_2}
    P(I) = \lambda_0^2 K \Bigl(\frac{\delta \omega}{\omega_{\rm Q}}\Bigr)^{4K}\sum_N \delta \omega \int dt e^{-i(\pi I/e - \omega_N)t}e^{-\gamma |t|}\sum_{m = 1}^N \frac{(4K \ln(\pi I / e\delta \omega) e^{-\gamma|t|})^{m-1}}{(m-1)!}.
\end{equation}
The sum over $m$ is dominated by a typical number of the emitted photons $m_{\rm typ}$ satisfying
\begin{equation}\label{eq:m_typ}
    m_{\rm typ}(I) \approx 4K \ln(\pi I / e\delta \omega).
\end{equation}
We see that although $m_{\rm typ} \gg 1$ is logarithmically large in parameter $\pi I / e\delta \omega\gg 1$ it is still relatively small compared to the relevant values of $N\approx \pi I / e\delta \omega$ in Eq.~(\ref{eq:P_sums_2}), i.e. $m_{\rm typ}\ll N$. This allows us to extend the summation over $m$ in Eq.~\eqref{eq:P_sums_2} to infinity. As a result, we find
\begin{align}\label{eq:P_sums_3}
    P(I) &= 
    \lambda_0^2 K \left(\frac{\pi |I|}{e\omega_{\rm Q}}\right)^{4K}\sum_N \delta \omega \int dt e^{-i(\pi I/e - \omega_N)t}e^{-\gamma |t|}e^{-4K \ln(\pi |I| / e\delta \omega)(1-e^{-\gamma |t|})},
\end{align}
where we wrote $(\delta \omega / \omega_Q)^{4K}$ as $(\pi |I|/e\omega_{\rm Q})^{4K} \cdot \exp\{-4K\ln(\pi |I|/e\delta \omega)\}$.
Lastly, we use~\footnote{In the left-hand side of Eq.~(\ref{eq:summation}), we extended the lower limit of summation to $-\infty$,
as $|N- \pi I/e| \ll \pi I/e$ for relevant $N$} identity 
\begin{equation}\label{eq:summation}
    \sum_N e^{i\omega_N t} = \frac{2\pi}{\delta \omega} \sum_k \delta (t - 2\pi k/\delta \omega)
\end{equation}
and the relation  $V(I)=P(I)/I$ to obtain: 
\begin{eqnarray}
&V(I) & = 2\pi \lambda_0^2 K\cdot\frac{1}{I}\left(\frac{\pi |I|}{e\omega_{\rm Q}}\right)^{4K} F\left({\pi I }/{e \delta \omega} \right), \qquad F(\Omega/\delta\omega)= \sum_k F_k (\Omega)  \cos\left(\frac{2\pi k \Omega}{\delta\omega} \right),
\label{eq:VI_resonance}\\
&  
F_k(\Omega) & =  \exp \left[-\frac{2\pi\gamma|k|} {\delta \omega} - 4K \ln(|\Omega| / \delta \omega)(1-e^{-2\pi\gamma |k| / \delta \omega})\right].
\label{eq:Fk}
\end{eqnarray}
Equations~(\ref{eq:VI_resonance}) and (\ref{eq:Fk}) comprise the main result of this Section. 

Setting $K\to 0$ in Eq.~(\ref{eq:Fk}), we recover from Eq.~(\ref{eq:VI_resonance}) the expected classical-limit result: a system of fixed-width Lorentzian resonances in $V(I)$. At finite $K$, the probability of ``splitting'' the Bloch oscillations quantum  into many emitted photons grows with $\Omega$. 
The enhanced width exceeds the natural one, $(e/\pi)\gamma$, at $4K\ln(\Omega/\delta\omega)\gg 1$. The latter condition allows us to expand the exponent in the parentheses in the Fourier coefficients (\ref{eq:Fk}) to linear order in $k$ before performing summation over $k$ in Eq.~(\ref{eq:VI_resonance}). On the other hand, as long as $4K\ln(\Omega/\delta\omega)\ll \delta\omega/\gamma$ and $\Omega$ is close to a resonant frequency, we are able to replace the summation by integration \footnote{Linearizing the exponential in Eq.~\eqref{eq:Fk} yields the Fourier series $F(x)=\sum_k e^{-k \gamma_N/d\omega}\cos(2\pi k x)$; compare with the Fourier series of $\theta_q^2=1+2\sum_k r^k \cos(2\pi k  \omega/\delta\omega)$. The change to integration is consistent with using Eq.~\eqref{eq:thetaq_lorentzians}}. The result is a sequence of resonances in $V(I)$ at $I= Ne\delta\omega/\pi$,
\begin{equation}\label{eq:well_resolved}
    V(I) = 2\pi \lambda_0^2 K\cdot\frac{1}{I}\Bigl(\frac{\pi |I|}{e\omega_{\rm Q}}\Bigr)^{4K} \sum_N  \frac{\delta \omega}{\pi} 
    \frac{\gamma_N}{(\pi I/e - N \delta \omega)^2 + \gamma_N^2},\qquad \gamma_N= \gamma(1 + 4 K \ln N )\,.
\end{equation}
The width of the $N$-th peak is proportional to the typical number of the emitted photons associated with it, $\gamma_N=\gamma\cdot m_{\rm typ}(N e \delta \omega / \pi)$, see Eq.~(\ref{eq:m_typ}). Equation~\eqref{eq:well_resolved} allows us to establish the range of currents for which our perturbative-in-$\lambda$ calculation holds: At a resonance, $V(I)\sim K  \lambda_0^2 \delta \omega / I \gamma$; yet, the voltage across the transmon may not exceed $\sim \lambda_0 / e$, leading to the condition $I\gg \lambda_{0} / e Z_2(0)$.

By comparing $\gamma_N$ with $\delta\omega$, we expect an exponentially large number $N_{\rm res}$ of resolved resonances in $V(I)$, 
\begin{equation}\label{eq:Nres}
\ln N_{\rm res}\sim\pi Z_1^2(0)/2R_Q Z_2(0) 
,\qquad 
I_{\rm res}=N_{\rm res}\frac{e\delta\omega}{\pi}\,.
\end{equation}
For higher currents, $I \gtrsim I_{\rm res}$, the resonant structure in the $VI$-relation is smeared out. 
We can find the $VI$-relation by noting that the  summation over $k$ in Eq.~\eqref{eq:Fk} converges rapidly in this regime. For $I \gg I_{\rm res}$, the leading contribution to $V(I)$ comes from terms with $k\in \{-1, 0, 1\}$. Discarding all other terms, we get
\begin{equation}\label{eq:VI_noresonance}
    V(I) = 2\pi \lambda_0^2 K\cdot\frac{1}{I}\Bigl(\frac{\pi |I|}{e\omega_{\rm Q}}\Bigr)^{4K} \Bigl(1 + 2 e^{-8\pi (K\gamma/\delta\omega) \ln(\Omega / \delta \omega)}\cos(2\pi \Omega / \delta \omega)\Bigr)\Bigr|_{\Omega = \pi I / e}.
\end{equation}
\end{widetext}
The sharp resonances present at $I\lesssim I_{\rm res}$ are washed out, and replaced by an exponentially small oscillatory term here. The leading monotonic dependence in $V(I)$ is insensitive to the mismatch and agrees~\footnote{At $K\ll 1$, the agreement is complete and includes the $K$ dependence of the prefactor in the right-hand side of Eq.~(\ref{eq:VI_noresonance}). An insignificant discrepancy appearing at $K\lesssim 1$ stems from the logarithmic approximation accepted in Eq.~(\ref{eq:logs}), which ignores a small probability of emitting more than one photon in the same mode. More precise accounting resolves the discrepancy \cite{RemezForthcoming}.} with the result for a junction attached to a semi-infinite superinductor of impedance $Z_1(0)$ \footnote{We find the voltage--current relation of a transmon coupled to a semi-infinite array in a complementary work \cite{shapiros_paper}.}.  
Forthcoming Sections focus on the weakly-inelastic regime of well-resolved resonances, $I \lesssim I_{\rm res}$. 

\begin{figure}[t]
    \centering
    \includegraphics[width = 8.6cm]{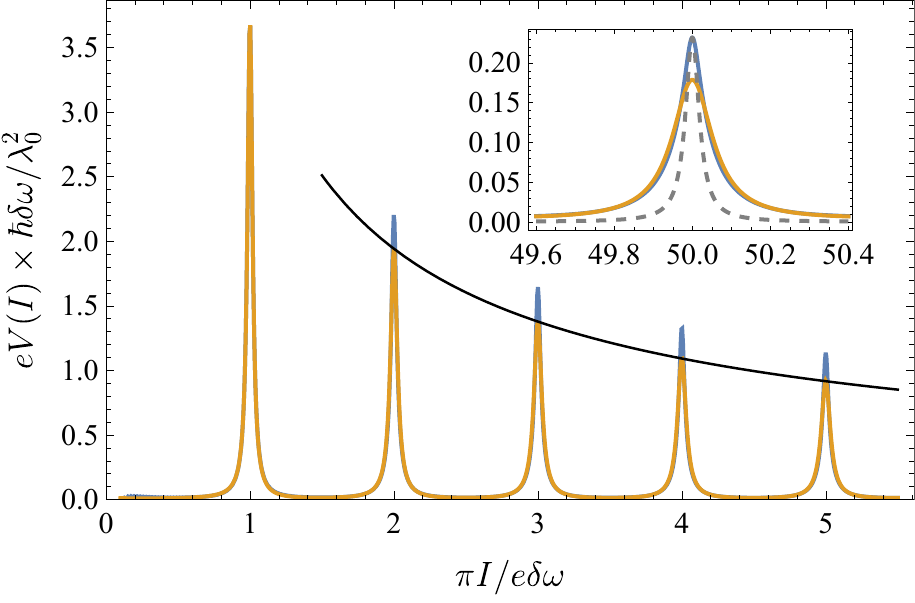}
    \caption{Voltage--current relation [Eq.~\eqref{eq:VI_resonance}, blue] and its Lorentzian approximation [Eq.~\eqref{eq:well_resolved}, orange].  The solid black line is $\sim I^{4K-1}/{[1 + 4K \ln (\pi I/ e \delta \omega)]}$. The decreasing peak amplitudes reflects the underlying insulating state of the junction.   Here we used $K=1/8$ and a large $Z_2 = 2~\mathrm{k\Omega}$ ($r \approx 0.86$; $\gamma/\delta\omega\approx0.2$) to broaden the resonances for clarity.  [Though Eqs.~\eqref{eq:VI_resonance} and \eqref{eq:well_resolved} were derived for large mode number, see inset, they  also capture low mode numbers qualitatively.] Inset: Closeup of the resonance at the 50th mode; the dashed curve is a reference Lorentzian with width $\gamma$. }
    \label{fig:fig2}
\end{figure}

To illustrate our results we will use experimentally-relevant parameters $Z_1 = 4 R_Q$ ($K=1/8$), $\delta \omega/2\pi = 100$~MHz, and $\omega_{\rm Q} / 2\pi = 10$~GHz throughout.. 
In Fig.~\ref{fig:fig2} we plot the voltage--current relation of Eq.~\eqref{eq:VI_resonance}, and its limit for well-resolved resonances \footnote{Assuming $E_C \approx 1$~GHz, $\lambda_0/e Z_2 < \delta \omega$ and our perturbative results extend down to the first mode}, Eq.~\eqref{eq:well_resolved}.  We note that even at $N\sim 1$ the latter gives a good approximation for the full result, Eq.~\eqref{eq:VI_resonance}.
Having established its validity, we will defer to this Lorentzian approximation in the next Sections. In the inset we show the logarithmic broadening of the resonances at higher mode numbers.

\section{Photon resonances in the Bloch oscillations radiation}
\label{sec:radiation}
{Due to the Bloch oscillations, the transmon radiates waves into its environment. In this section, we find the radiation spectrum $S(\omega, \Omega)$ of a transmon biased by current $I \equiv e \Omega / \pi$, see Eq.~\eqref{eq:bloch_freq}; $S(\omega, \Omega)$ is a number of photons emitted in a frequency interval $[\omega, \omega + d\omega]$ per unit time.}

Applying the Fermi golden rule to evaluate the time derivative of a photon mode's occupation, we derive 
\begin{equation}\label{eq:Somega}
    S(\omega,\Omega) = \frac{K\lambda_0^2}{\omega}{\cal C}_\theta(\Omega - \omega)\sum_{n = 0}^{\infty} \frac{\delta \omega}{\pi}\frac{ \gamma}{\gamma^2 + (\omega - n\delta\omega)^2},
\end{equation}
in close analogy with the derivation of $P(I)$.   ${\cal C}_\theta(\Omega-\omega)$ is the same correlation function as in Eq.~(\ref{eq:P_to_C}). Using the relations between $V(I)$, $P(I)$, and ${\cal C}_\theta (\Omega)$ in Eqs.~(\ref{eq:P_to_C}) and (\ref{eq:VI_resonance}), it is easy to recover ${\cal C}_\theta(\Omega-\omega)$ in the form 
\begin{eqnarray}
&&{\cal C}_\theta(\Omega^\prime)= \frac{8\pi K}{\omega_Q}\Bigl(\frac{\Omega^\prime}{\omega_{\rm Q}}\Bigr)^{4K-1} F(\Omega^\prime/\delta\omega)\Theta(\Omega^\prime),
\label{eq:Ctheta}
\end{eqnarray}
with $F(\Omega/\delta\omega)$ given by Eq.~(\ref{eq:VI_resonance}).

\begin{figure}
    \centering
    \includegraphics[width = 8.6cm]{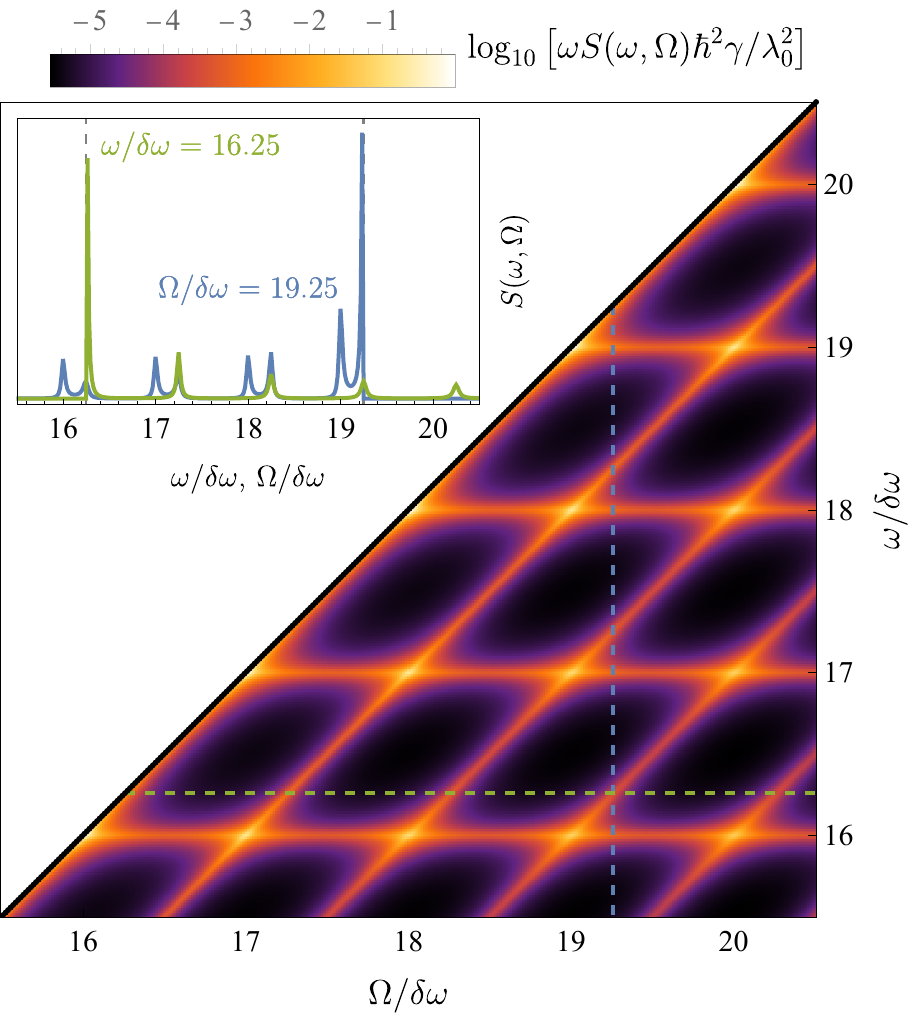}
    \caption{Emission power spectral density $\omega S(\omega,\Omega)$ [Eq.~\eqref{eq:Somega_well_resolved}] as a function of  Bloch oscillation frequency $\Omega$ and photon frequency $\omega < \Omega$. Emission is peaked at waveguide mode frequencies $\omega_n$ and the drive frequency $\Omega$, and is maximal when these coincide. Dashed lines indicate the cross sections in the inset. Inset: Slices at fixed $\Omega$ and $\omega$. 
    For fixed $\Omega$ (blue) away from a resonance, emission peaks appear in pairs, corresponding to excitations of photons at  $
    \omega = \omega_n$ and `evanescent' photons at $\omega = \Omega - \omega_{n'} $. Their heights are $\sim\mathrm{constant}$ and $\sim(\Omega - \omega)^{4K-1}$, respectively,  and therefore the former peaks have comparable height while the latter grow as $\omega$ approaches $\Omega$  [cf. Eq.~\eqref{eq:Somega_well_resolved}]. 
    At low $\omega$ only the former survive. 
    For fixed $\omega$ (green), the emission power peak amplitudes increase  increases with $\Omega>\omega$. Circuit parameters are the same as in Fig.~\ref{fig:fig2}}
    \label{fig:fig3}
\end{figure}

As discussed above, Bloch oscillations of a fixed frequency $\Omega$ excite multiple photons, whose energies sum up to $\Omega$. 
Thus, clearly $S(\omega,\Omega)\neq 0$ only at $\omega<\Omega$; this is reflected by the step function in ${\cal C}_\theta(\Omega^\prime)$
in Eq.~(\ref{eq:Ctheta}). At a moderate inelasticity, $4K\ln[(\Omega-\omega)/\delta\omega]\ll\delta\omega/\gamma$, function  $F$ exhibits resonant structure [given by the sum in Eq.~(\ref{eq:well_resolved})]. Therefore, the radiation spectrum $S$ as a function of $\omega$ and {$I = \pi \Omega / e$} has resonances in both variables:
\begin{align}
       & S(\omega, \Omega)  =   \frac{8 \pi K^2 \lambda_0^2}{\omega_{\rm Q}} \frac{1}{ \omega } \Bigl(\frac{\Omega - \omega}{\omega_{\rm Q}}\Bigr)^{4K-1}
       \times \nonumber \\ &
       \sum_{n = 0}^{\infty} \frac{\delta \omega}{\pi}\frac{ \gamma}{\gamma^2 + (\omega - n\delta\omega)^2}
    \sum_{N=0}^{\infty} \frac{\delta \omega}{\pi} 
    \frac{\gamma_N \cdot \Theta(\Omega - \omega)}{(\Omega - \omega- N\delta\omega)^2 + \gamma_N^2}. \label{eq:Somega_well_resolved}
\end{align} 
The total emission power is the largest when $\Omega$ hits a resonance with one of the modes in the array. When this occurs, emission $S(\omega, \Omega)$ has sharp peaks at $\omega = \omega_n$ with $\omega_n < \Omega$. Here the resonances in $\Omega$ are of the same nature as those in $V(I)$ dependence.  
The power emission spectrum $\omega \times S(\omega, \Omega)$ is illustrated in Fig.~\ref{fig:fig3}.

\section{Up-mixing of Bloch oscillations by a Josephson plasmon}
\label{sec:upmixing}
Bloch oscillations in the current-biased transmon also affect the transmon's microwave response properties, such as its fluorescence spectrum. In this Section, we find the cross-section $\sigma(\omega \rightarrow \omega^\prime)$ of inelastic scattering of a microwave photon off the transmon. We show that mixing of the photon with Bloch oscillations creates thresholds in $\sigma(\omega \rightarrow \omega^\prime)$ at $\omega^\prime = \omega \pm \Omega$ (as in the previous Section, $\Omega = \pi I / e$ is the Bloch oscillations frequency). The threshold behavior is determined by the line impedance $Z_1(0)$, and further modified by the resonances in the environment. 

Specifically, we consider how a photon is scattered by the transmon in nearly resonant conditions, i.e., when the photon frequency $\omega$ is close to the transmon frequency $\omega_{\rm Q}$. Such a scattering occurs predominantly in the elastic channel. However, with a small-yet-nonzero probability the scattering may be accompanied by a phase slip. This makes scattering inelastic and dependent on the quantum dynamics of the transmon. To describe this inelastic scattering, we use the effective Hamiltonian derived in Ref.~\cite{Houzet2020}:
\begin{align}\label{eq:eff_ham_conv}
    &H_{\rm eff} = \omega_{\rm Q} |1\rangle\langle 1| + \!\!\!\sum_{\omega_q > \omega_{\rm c}} \omega_q |q\rangle \langle q| -i \sum_q t_q \bigl(|1\rangle \langle q| - |q\rangle\langle 1|\bigr)\notag\\
    &+ \!\!\!\!\sum_{0<\omega_q<\omega_{\rm c}}\!\!\!\!\omega_q a_q^\dagger a_q + \lambda_1 |1\rangle\langle 1| \cos (2\tilde{\theta}(x=0)\! - \!\Omega t)
\end{align}
(we remind one that $\Omega = \pi I / e$, see Eq.~\eqref{eq:bloch_freq}). The first line here describes the coupling of high-frequency ($\omega \sim \omega_{\rm Q}$) photons in the junctions array to the transmon ($|1\rangle$ is the the first excited state of the transmon). The matrix elements $t_q$ contain information about resonance of $|0\rangle\rightarrow |1\rangle$ transition of the transmon with the array modes:
\begin{equation}
    |t_q|^2 = \frac{\Gamma \Delta}{\pi} \sum_n \frac{\gamma \delta \omega / \pi}{(\omega_q - n \delta \omega)^2 + \gamma^2}.
\end{equation}
Here $\Delta = \pi v_2 / L$ is the frequency spacing of the modes in the entire environment; $1/\Gamma$ would be the radiative lifetime of the excited state $|1\rangle$ of the transmon were it connected to a semi-infinite array ($x_0 \rightarrow \infty$): 
\begin{equation} 
    \Gamma = \frac{1}{2Z_1(0)C}.
\end{equation}
Here $C$ is the capacitance of the junction related to its charging energy $E_C = e^2 / 2C$. The excited state $|1\rangle$ decays because of the transmon hybridization with the array, therefore $1/\Gamma$ scales with the impedance $Z_1(0)$. 

In the following, we consider the most interesting case~\cite{Kuzmin2019a} of a relatively strong hybridization, 
\begin{equation}\label{eq:strongh}
\Gamma\gg\delta\omega.
\end{equation}
In this limit, hybridization shifts frequencies $\omega_n$ of a large number of the superinductor eigenmodes with $\omega_n$ within the interval $|\omega_n-\omega_Q|\lesssim\Gamma$. The shifted frequencies $\omega_n$ are solutions~\cite{Fano1961,Houzet2020} of the equation 
\begin{equation} \label{eq:fano_freqs}
\omega_n - \omega_{\rm Q} = \Gamma \cot (\omega_n / \delta\omega).
\end{equation}
(The waveguide plasmon spectrum is generally dispersive, so the ``bare'' mode spacing around $\omega_Q$ will be a somewhat smaller $\delta\omega'<\delta\omega$. The hybridization in Eq.~\eqref{eq:fano_freqs} can already produce comparable densification \cite[see e.g.~Ref.~][]{Mehta2023}. For clarity, we neglect dispersion, but our results below may be easily generalized.)
The low-impedance external circuit broadens these levels into narrow resonances of width $\gamma\ll\delta\omega$. 

The second line of Eq.~\eqref{eq:eff_ham_conv} describes coupling of state $|1\rangle$ to the low-frequency photons via phase slips; the phase slip amplitude $\lambda_1$ is given by Eq.~(\ref{eq:lambdas}).
In the displacement charge $\tilde{n}$, we account only for the low-frequency modes, $0 < \omega < \omega_c \equiv v_2 q_c$:
\begin{equation}\label{eq:tildetheta}
    \tilde{\theta}(x=0) = \sum_{0 < q < q_c} \sqrt{\frac{K \Delta}{\omega_q}}\theta_q \bigl(a_q + a_q^\dagger\bigr),
\end{equation}
cf. Eqs.~(\ref{eq:theta}) and (\ref{eq:theta_q_mis}). Scale $\omega_{\rm c}$ separates the low- and high-frequency modes, $\Gamma_{\rm Q} \ll \omega_{\rm Q} - \omega_{\rm c}, \omega_{\rm c} \lesssim \omega_{\rm Q}$. The final results are independent of a specific value $\omega_{\rm c}$, as long as $K<1/2$, see Ref.~\cite{Houzet2020}.

We begin by diagonalizing the first line of Eq.~\eqref{eq:eff_ham_conv} and expanding state $|1\rangle$ in eigenstates $|k\rangle$,
\begin{equation}\label{eq:beta}
    |\beta_k|^2 = |\langle k | 1\rangle|^2 = \sum_{n = 0}^{+\infty}
    \frac{\Gamma\delta \omega/\pi}{(\omega_{\rm Q} - \omega_n)^2 + \Gamma^2}
     \frac{\gamma\Delta/\pi}{(\omega_k - \omega_n)^2+\gamma^2}. 
\end{equation}
In terms of the hybridized states, the Hamiltonian acquires the form
\begin{align}\label{eq:eff_ham_conv_2}
    H_{\rm eff} &= \sum_{k > q_c} \omega_k |k\rangle \langle k| +\hspace{-0.2cm} \sum_{0 < q < q_c} \omega_q a_q^\dagger a_q +V_1 e^{-i\Omega t} + V_1^\dagger e^{i\Omega t},\notag\\
    V_1 &= \frac{\lambda_1}{2} \exp(2\pi i\,\tilde{n}) \sum_{k,k^\prime} \beta_k \beta^\star_{k^\prime} |k\rangle \langle k^\prime|.
\end{align}
Let us now assume that the system is initially in a state with a single photon of energy $\omega_k$:  $|i\rangle = |k, 0\rangle$ ($0$ indicates that there are no ``soft'' photons in the line). We are interested in the inelastic scattering cross-section of this photon into a photon with a different (yet close) frequency $\omega_{k^\prime}$. The respective final state is $|k^\prime, f\rangle$, where $f$ abbreviates the multiphoton state formed by a number of ``soft'' photons.
The inelastic scattering may reduce or increase the frequency of the outgoing photon. The former process happens already at zero bias. In contrast, inelastic scattering to higher frequency occurs exclusively due to the up-mixing of $\Omega$ with the incoming photon of frequency $\omega_k$. In the following, we will focus only on the latter, anti-Stokes component of inelastic scattering.

We can evaluate the scattering cross-section   $\sigma^{\phantom\dagger}_{\rm aS}(\omega_k~\rightarrow~\omega_{k^\prime})$  for this process using Fermi's golden rule:
\begin{align}\label{eq:cross_sec_sas}
    \sigma^{\phantom\dagger}_{\rm aS}& (\omega_k \rightarrow \omega_{k^\prime}) =\notag\\ \frac{{4\pi^2}}{\Delta^2} &\sum_{f} |\langle k^\prime, f| V_1 | k, 0\rangle|^2 \delta (\omega_{k^\prime} + E_f - (\omega_k + \Omega)).
\end{align} 
Using Eqs.~\eqref{eq:beta} and \eqref{eq:eff_ham_conv_2}, we can rewrite Eq.~\eqref{eq:cross_sec_sas} as
\begin{align}\label{eq:stokes}
    &\sigma^{\phantom\dagger}_{\rm aS} (\omega \rightarrow \omega^\prime) = \frac{1}{2\pi}\frac{\Gamma^2 \lambda_1^2\, \widetilde{{\cal C}}_\theta(\Omega + \omega -\omega^\prime)}{[(\omega_{\rm Q} - \omega)^2 + \Gamma^2][(\omega_{\rm Q} - \omega^\prime)^2 + \Gamma^2]}
    \notag\\
   &\times \sum_{n,l = 0}^{+\infty}
\frac{\gamma\delta\omega/\pi}{(\omega - \omega_n)^2+\gamma^2}  \frac{\gamma\delta\omega/\pi}{(\omega^\prime - \omega_l)^2+\gamma^2}\,.
\end{align}
Here the correlation function $\widetilde{{\cal C}}_\theta(\Omega)$ is defined the same way as ${\cal C}_\theta(\Omega)$ in Eq.~(\ref{eq:P_start}), with replacement $\theta(x=0)\to\tilde{\theta}(x=0)$, see Eqs.~(\ref{eq:theta}) and (\ref{eq:tildetheta}). This amounts to substituting $\omega_{\rm Q} \to \omega_c$ in Eq.~\eqref{eq:Ctheta}; since $\mathcal{C}_\theta$ is only weakly dependent on the cutoff and $\omega_c$ is of the order of $\omega_{\rm Q}$, at our level of treatment we may continue to use Eq.~\eqref{eq:Ctheta} as it appears. Equation~(\ref{eq:stokes}) is the main general result of this Section.  It allows one to evaluate the anti-Stokes component of inelastic scattering for arbitrary frequencies of the absorbed and emitted photons, and arbitrary Bloch frequency. 

\begin{figure*}[tb]
    \centering
    \includegraphics[width=8.6cm]{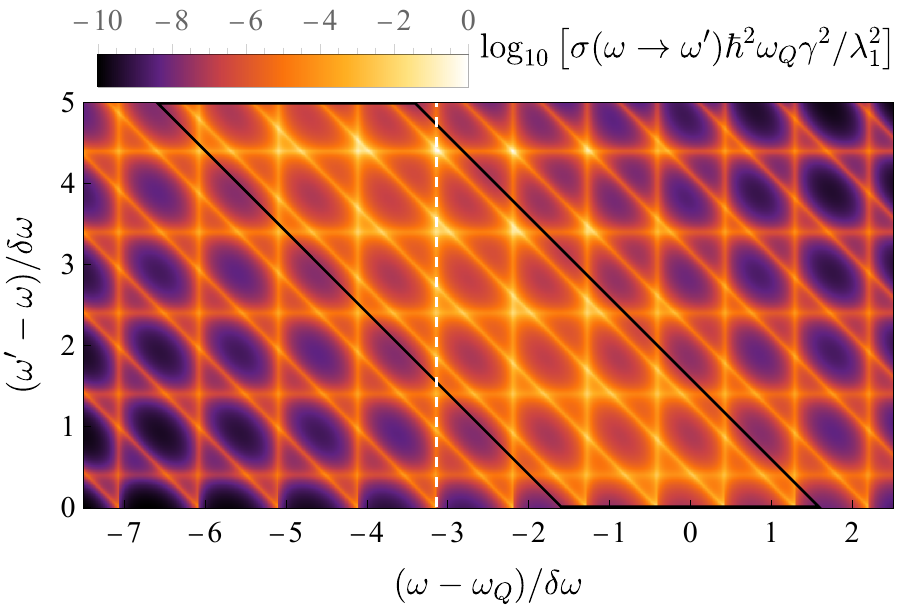}\hfill
    \includegraphics[width=8.6cm]{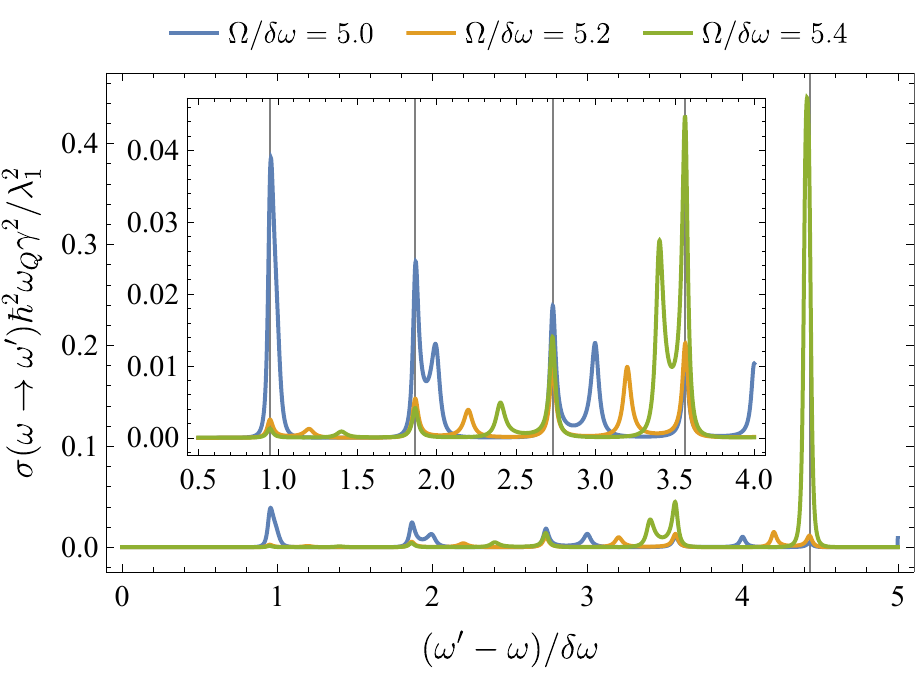}
    \caption{Bloch oscillation upmixing by transmon flourescence. (a) Anti-Stokes scattering cross section $\sigma_{\mathrm{aS}}(\omega\to\omega')$ [Eq.~\eqref{eq:resonant-stokes}] at fixed $\Omega=5.4\delta\omega$; note the vertical axis is the relative $\omega'-\omega$. The black diamond indicates the region where $|\omega-\omega_Q|,|\omega'-\omega_Q|\apprle\Gamma$. Vertical, diagonal, and horizontal streaks correspond to the resonance conditions $\omega = \omega_{l}$, $\omega'=\omega_l$, and $\Omega+
    \omega-\omega'=n\delta\omega$ [cf. Eqs.~\eqref{eq:fano_freqs} and \eqref{eq:resonant-stokes}]. The latter are swept upwards with increasing $\Omega$; as the hybridized modes are not equidistant (note 11 lines fit in a span of $10 \delta \omega$), general values of $\Omega$ would lead to some triple-intersection somewhere in the $(\omega,\omega')$ plane where upconversion is strongest, here around $(\omega-\omega_Q,\omega'-\omega)/\delta\omega \approx (-3,4)$. 
    (b) Upconversion spectrum for fixed incoming $\omega$ corresponding to the dashed white line in (a), for different $\Omega$. Relatively minor changes in $\Omega$ can have dramatic effects on the upconversion spectrum, as the resonance condition for $\Omega+\omega-\omega'$ moves in and out of alignment with that for $\omega,\omega'$. Here the emission lines switch from decreasing with $\omega'$ at $\Omega = 5.0 \delta\omega$ [blue] to  increasing  with $\omega'$ at $\Omega=5.4 \delta \omega$ [green, same as in (a)]. This modulation with $\Omega$ is approximately periodic with period $\delta\omega$. The vertical lines indicate the positions of the hybridized modes [solutions to Eq.~\eqref{eq:fano_freqs}]. We use the same circuit parameters as in Figs.~\ref{fig:fig2} and \ref{fig:fig3}.}
    \label{fig:fig4}
\end{figure*}

Again assuming weak inelasticity $4K\ln(\Omega/\delta\omega)\ll\delta\omega/\gamma$, we approximate $F$ by a sequence of Lorentzians.
In addition, we will assume that the incoming photon frequency is in the vicinity of one of the ``environmental'' resonances $\omega_{n_0}$ close to the transmon frequency, $|\omega_{n_0}-\omega_{\rm Q}| \ll \Gamma$. These simplifications allow us to keep in Eq.~(\ref{eq:stokes}) only one term in the sum over $n$, and to use the counterpart of Eq.~(\ref{eq:well_resolved}) for $\widetilde{{\cal C}}_\theta(\Omega^\prime)$,
\begin{align}\label{eq:resonant-stokes}
    &\sigma^{\phantom\dagger}_{\rm aS} (\omega \rightarrow \omega^\prime) = \frac{4\lambda_1^2}{(\omega - \omega_{n_0})^2+\gamma^2}\cdot\frac{(\gamma/\pi)^2}{(\omega^\prime-\omega_{\rm Q})^2 + \Gamma^2}
    \notag\\
   &\times 
     \sum_{l = 0}^{+\infty} \frac{\delta\omega^2}{(\omega^\prime - \omega_l)^2+\gamma^2}\cdot\frac{K}{\omega_Q}\Bigl(\frac{\Omega +\omega-\omega^\prime}{\omega_{\rm Q}}\Bigr)^{4K-1}\\
    & \times\frac{\delta \omega}{\pi} \sum_{N=0}^{\infty} 
    \frac{\gamma_N \Theta(\Omega + \omega -\omega')}{(\Omega +\omega-\omega^\prime - N \delta \omega)^2 + \gamma_N^2}.
    \nonumber
\end{align}
Here $\gamma_N$ is given in Eq.~(\ref{eq:well_resolved}).
It is clear from Eq.~(\ref{eq:resonant-stokes}) that the cross-section $\sigma^{\phantom\dagger}_{\rm aS} (\omega \rightarrow \omega^\prime)$ exhibits a resonant structure as a function of the incoming and outgoing photon energies $\omega,\omega^\prime$, and the energy transfer $\Omega+\omega-\omega'$. 
We illustrate this  in Fig.~\ref{fig:fig4}. In the first panel we show $\sigma_{\rm a S} (\omega \to \omega')$ at fixed Bloch frequency. Upmixing is expectedly most prominent when the incoming and outgoing photons are within $\Gamma$ of $\omega_{\rm Q}$. The process is further enhanced for each satisfied resonant condition, that is when $\omega$, $\omega'$, and $\Omega+\omega-\omega'$ each coincide with waveguide modes. Maximal upmixing efficiency occurs at ``triple-resonance'',
though generically there will be some tension in  satisfying all three resonance conditions simultaneously. 
As the hybridization with the transmon densifies the modes around $\omega_{\rm Q}$ \cite{Mehta2023}, for resonant $\omega,\omega'$ this occurs at $\Omega$ detuned somewhat below the low-frequency modes $n \delta \omega$ (waveguide dispersion will further magnify this detuning). Therefore, for fixed and resonant photon frequencies $\sigma_{\rm aS}$ attains maxima at $\Omega$ lesser than those which maximize $V(I)$. The hybridization additionally leads to nonuniform spacing of the high-frequency photon modes. This increases the likelihood that at fixed and  general values of $\Omega$, some resonant pair $(\omega,\omega')$ may achieve triple-resonance. The proximity to this triple-resonance affects $\sigma_\mathrm{aS}$ much more strongly than the $(\Omega+\omega-\omega')^{4K-1}$ dependence in Eq.~\eqref{eq:resonant-stokes}; therefore, the resonance peaks of $\sigma_\mathrm{aS}$ might not be monotonically increasing as  $\omega'$ approaches $\Omega+\omega$. This is illustrated in the second panel.

\section{Conclusions} \label{sec:conclusions}

In this work we describe Bloch oscillations in a small Josephson junction embedded in a high-impedance electromagnetic environment. That such environments can be implemented by Josephson junction arrays, with plasmon standing wave modes, begs the question of their  importance. Indeed, the interaction between these photons and the (unbiased) junction  is well understood \cite{Burshtein2021, Mehta2022,Mehta2023, Burshtein2023}. Our goal here was to extend this understanding to the reciprocal influence of environmental resonances on Bloch oscillations, and vice versa.  We make predictions for the voltage--current relation, radiation spectrum, and inelastic scattering rate.

Let us recapitulate our main results and their experimental implications. We found that resonant peaks appear in the $VI$ relation when the Bloch oscillation frequency $\Omega = \pi I / e$ due to a DC bias coincides with a waveguide mode frequency $\omega_n$. Multiphoton processes broaden the resonances, as shown in Eq.~(\ref{eq:well_resolved}). The broadening increases with the current, and eventually the multiphoton processes wash out the resonant structure, see~Eq.~\eqref{eq:VI_noresonance}.

Besides direct measurement of the $VI$ characteristic, Bloch resonances can be evidenced in the radiation spectrum emitted to the transmission line. The photon emission rate, given by Eq.~\eqref{eq:Somega_well_resolved}, shows resonance in both the photon frequency $\omega$ and Bloch frequency $\Omega$. A possible experiment would measure the emission at fixed frequency $\omega$ as the bias current is increased. The signature of Bloch oscillations would be a lack of emission until $I > e \omega / \pi$, with subsequent periodic modulation of the emission intensity with period $\Delta I = e \delta \omega /\pi$.

Bias currents in the picoampere range \cite{KuzminMarchMeeting2023}  will lead to MHz-range emitted photons, which may be hard to detect. We thus consider inelastic scattering of photons near the qubit transition frequency $\omega_{\rm Q}$. In the absence of current, an incoming photon may ``split'' in a number of photons of lower frequency, leading to the Stokes component of inelastic scattering observed in Refs.~\cite{Kuzmin2021, Kuzmin2023}. Bloch oscillations result in the possibility of blue-shifting the incoming photon to higher frequency. We predict the appearance of the anti-Stokes inelastic scattering component, and elucidate the structure of the respective scattering cross-section $\sigma(\omega \to \omega')$, see Eq.~\eqref{eq:resonant-stokes}. For an incoming photon of frequency $\omega=\omega_{\rm Q}$, the anti-Stokes side-band extends to $\omega' =\omega_{\rm Q} + \Omega$. Once more, the scattering process is resonant for energy exchange $\Omega + \omega - \omega'$ coinciding with waveguide modes. This constraint may be frustrated due to photon nonlinear dispersion in the junction array, which is outside the scope of this work.

We may compare our results to recent experiments  \cite{Kuzmin2022, Mehta2023,Kuzmin2023}. In these devices, impedances as high as $Z_1(0)\approx 3R_Q$ are achieved, whereas $Z_2(0) \sim 50~\Omega$. The Josephson junction array is typically a few millimeters long, with low-frequency wave velocity $v\sim 10^6 ~\mathrm{m/s}$, so that $\delta \omega/2\pi \sim 100~\mathrm{MHz}$. Therefore $\sim 10^2$ modes are present below the cut-off plasma frequency $\Omega_P$ of the transmission line; typically $\Omega_P/2\pi  \sim  20 ~\mathrm{GHz}$. Under these circumstances, $\ln N_\mathrm{res} \apprge 10^3$ [Eq.~\eqref{eq:Nres}], and  all frequencies below $\omega_{\rm Q} \sim $ 1--10 GHz are in the weakly-inelastic regime. We then expect well-separated resonances given by Eq.~\eqref{eq:VI_resonance}.

Circuits comprised of a small Josephson junction galvanically coupled to a high-impedance transmission line were mostly probed by microwave spectroscopy, which revealed the presence of inelastic scattering channel, along with the elastic one \cite{Kuzmin2021,Kuzmin2023}. In the latest work of that cycle, an auxiliary measurement of the DC voltage-current characteristic was also performed \cite{KuzminMarchMeeting2023}. The found non-monotonic $VI$ curve carried traits consistent with the presence of Bloch oscillations. However, there was no clear observation of the resonances associated with the discrete modes of the transmission line. Further experiments on this class of circuits, focusing on the signatures we outlined here --- $VI$ modulations, Bloch oscillation radiation spectrum, and anti-Stokes inelastic scattering --- will shed more light on this system.

We thank V.~Manucharyan, R.~Kuzmin, M.~Houzet, M.~Goldstein, and A.~Burshtein for fruitful discussions. This research was sponsored by the Army Research Office (ARO) under grant number W911NF-22-1-0053, by the Office of Naval Research (ONR) under award number N00014-22-
1-2764, by the NSF Grant No. DMR-2002275. B.~R. gratefully acknowledges support of the Yale Prize Fellowship.

\bibliography{references, aux_references}

\begin{thebibliography}{45}
\bibitem [{Ind()}]{Indices}%
  \BibitemOpen
  \href@noop {} {}\bibinfo {note} {We will use indices $k,q$ to enumerate
  photon eigenstates supported by the entire environment, and indices $n,l,N$
  to enumerate the discrete standing wave modes of the high-impedance
  array.}\BibitemShut {Stop}%
\bibitem [{J0()}]{J0}%
  \BibitemOpen
  \href@noop {} {}\bibinfo {note} {At strong inlasticity $K\sim 1$, $J_0(t)$
  can become important \cite{RemezForthcoming}.}\BibitemShut {Stop}%
\end{thebibliography}

\end{document}